\documentclass[conference,a4paper]{IEEEtran}

\usepackage{amssymb, amsmath, color}
\usepackage{amsthm,graphicx,enumerate,verbatim,xcolor}
\usepackage[small]{caption}
\usepackage{subfigure}
\usepackage{tikz}
\usepackage{bbm}
\usepackage{mathtools}

\newtheorem{lem}{Lemma}
\newtheorem{defn}{Definition}

\newtheorem{property}{Property}
\theoremstyle{remark}

\def\squarebox#1{\hbox to #1{\hfill\vbox to #1{\vfill}}}
\newcommand{\inout}{-}
\newcommand{\Pbar}{\overline{P}}

\newcommand{\eps}{\varepsilon}

\newcommand{\Ebb}{\mathbb{E}}

\newcommand{\Pbb}{\mathbb{P}}
\newcommand{\Rbb}{\mathbb{R}}

\newcommand{\Ccal}{\mathcal{C}}

\newcommand{\Vcal}{\mathcal{V}}

\newcommand{\Xcal}{\mathcal{X}}
\newcommand{\Ycal}{\mathcal{Y}}

\newcommand{\Lcal}{\mathcal{L}}
\newcommand{\cn}{\mathcal{C}^n}
\newcommand{\Pbf}{\mathbf{P}}
\newcommand{\Qbf}{\mathbf{Q}}

\newcommand\blfootnote[1]{%
  \begingroup
  \renewcommand\thefootnote{}\footnote{#1}%
  \addtocounter{footnote}{-1}%
  \endgroup
}

\begin{document}

\sloppy

%% Paper Title
%% You can use linebreaks \\ within to get better formatting as
%% desired. 
\title{The Likelihood Encoder for Lossy Source Compression}

%% Author names and affiliations:

\author{Eva C. Song \qquad Paul Cuff \qquad H. Vincent  Poor\\ Dept. of Electrical Eng., Princeton University,  NJ 08544\\ \{csong, cuff, poor\}@princeton.edu}

%% To balance the two columns, you should reduce the text-height of
%% the last page using the following command:
%%%%%%%%%%%%%%%%%%%%%%%%%%%%%%%%%%%%%%%%%%%%%%%%%%%%%%%%%%%%%%%%%%%%%
%\addtolength{\textheight}{-9.35cm}
%%%%%%%%%%%%%%%%%%%%%%%%%%%%%%%%%%%%%%%%%%%%%%%%%%%%%%%%%%%%%%%%%%%%%
%% with an appropriate value. This command must be place on the second
%% last page, i.e., for a one-page abstract here, for a two-page
%% abstract right after the \maketitle command.

%% Create the title:
\maketitle
\blfootnote{This research was supported in part by the Air Force Office of Scientific Research under Grant FA9550-12-1-0196 and MURI Grant FA9550-09-05086 and in part by National Science Foundation under Grants CCF-1116013 and CNS-09-05086.}
\begin{abstract}
%To be considered for an IEEE Jack Keil Wolf ISIT Student Paper Award.
In this work, a likelihood encoder is studied in the context of lossy source compression. The analysis of the likelihood encoder is based on a soft-covering lemma. It is demonstrated that the use of a likelihood encoder together with the soft-covering lemma gives alternative achievability proofs for classical source coding problems. The case of the rate-distortion function with side information at the decoder (i.e. the Wyner-Ziv problem) is carefully examined and an application of the likelihood encoder to the multi-terminal source coding inner bound (i.e. the Berger-Tung region) is outlined.
\end{abstract}

\section{Introduction} \label{intro}
Rate-distortion theory, founded by Shannon in \cite{shannon-math} and \cite{shannon-rd}, provides the fundamental limits of lossy source compression. The minimum rate required to represent an independent and identically distributed (i.i.d.) source sequence under a given tolerance of distortion is given by the rate-distortion function. Related problems such as source coding with side information available only at the decoder \cite{wz} and distributed source coding \cite{tung}, \cite{berger1977}, \cite{berger1989} have also been heavily studied in the past decades. Standard proofs \cite{cover}, \cite{network-it} of achievability for these rate-distortion problems often use joint-typicality encoding, i.e. the encoder looks for a codeword that is jointly typical with the source sequence. The distortion analysis involves bounding several ``error" events which may come from either encoding or decoding. These bounds use the joint asymptotic equipartition principle (J-AEP) and its immediate consequences as the main tool. In the cases where there are multiple information sources, such as side information at the decoder, intricacies arise, such as the need for a Markov lemma \cite{cover} and \cite{network-it}. These subtleties also lead to error-prone proofs involving the analysis of error caused by random binning, which have been pointed out in several existing works \cite{hybrid} \cite{lapidoth}.

In this paper, we propose using a likelihood encoder to achieve classical source coding results such as the Wyner-Ziv rate-distortion function and Berger-Tung inner bound. This encoder has been used in \cite{cuff-itw2013} to achieve the rate-distortion function for point-to-point communication and in \cite{cuff2012distributed} and \cite{cuff-permuter} to achieve strong coordination. The advantage of the likelihood encoder over a joint-typicality encoder becomes crucial in secrecy systems \cite{schieler-journal}.

Just as the joint-typicality encoder relies on the J-AEP, the likelihood encoder relies on the soft-covering lemma. The idea of soft-covering was first introduced in \cite{wyner} and was later used in \cite{han-verdu} for channel resolvability.

The application of the likelihood encoder together with the soft-covering lemma is not limited to only discrete alphabet. The proof for sources from continuous alphabets is readily included, %Unlike the type-covering lemma, which is only applicable to discrete alphabet, 
since the soft-covering lemma imposes no restriction on alphabet size. Therefore, no extra work, i.e. quantization of the source, is needed to extend the standard proof for discrete sources to continuous sources as in \cite{network-it}. This advantage becomes more desirable for the multi-terminal case, since generalization of the type-covering lemma and the Markov lemma to continuous alphabets is non-trivial. Strong versions of the Markov lemma on finite alphabets that can prove the Berger-Tung inner bound can be found in \cite{network-it} and \cite{coord}. However, generalization to the continuous alphabets is still an ongoing research topic. Some work, such as \cite{jeon}, has been dedicated to making this transition, yet is not strong enough to be applied to the Berger-Tung case.

\section{Preliminaries} \label{prelim}
\subsection{Notation} \label{notation}
A sequence $X_1,..., X_n$ is denoted by $X^n$. Limits taken with respect to ``$n\rightarrow \infty$" are abbreviated as ``$\rightarrow_n$". Inequalities with $\limsup_{n\rightarrow \infty}h_n\leq h$ and $\liminf_{n\rightarrow \infty}h_n\geq h$ are abbreviated as $h_n\leq_n h$ and $h_n\geq_n h$, respectively. %Throughout this paper, random variables are assumed to take only countably many values. 
When $X$ denotes a random variable, $x$ is used to denote a realization, $\mathcal{X}$ is used to denote the support of that random variable, and $\Delta_{\Xcal}$ is used to denote the probability simplex of distributions with alphabet $\Xcal$. The symbol $|\cdot|$ is used to denote the cardinality. A Markov relation is denoted by the symbol $\inout$. %For an i.i.d. sequence pair $(X^n,Y^n)$ distributed according to $\prod_{i=1}^n \Pbar_{XY}$, the joint distribution is denoted by $\Pbar_{X^nY^n}$ and the corresponding conditional probability $\prod_{i=1}^n\Pbar_{Y|X}$ is denoted by $\Pbar_{Y^n|X^n}$. 
We use $\Ebb_P$, $\Pbb_P$, and $I_{P}(X;Y)$ to indicate expectation, probability, and mutual information taken with respect to a distribution $P$; however, when the distribution is clear from the context, the subscript will be omitted. To keep the notation uncluttered, the arguments of a distribution are sometimes omitted when the arguments' symbols match the subscripts of the distribution, e.g. $P_{X|Y}(x|y)=P_{X|Y}$. We use a bold capital letter $\mathbf{P}$ to denote that a distribution $P$ is random. We use $\Rbb$ to denote the set of real numbers and $\Rbb^+$ to denote the nonnegative subset. 

For a distortion measure $d: \mathcal{X} \times \mathcal{Y}\mapsto \mathbb{R}^+$, we use $\Ebb \left[d(X,Y)\right]$ to measure the distortion of $X$ incurred by representing it as $Y$. The maximum distortion is defined as
$$d_{max}=\max_{(x,y) \in \Xcal\times\Ycal} d(x,y).$$
The distortion between two sequences is defined to be the per-letter average distortion 
$$d(x^n,y^n)=\frac1n\sum_{t=1}^n d(x_t,y_t).$$

\subsection{Total Variation Distance}
The total variation distance between two distributions $P$ and $Q$ on the same alphabet $\Xcal$ is defined as
$$\lVert P-Q\rVert_{TV}\triangleq \sup_{A}|P(A)-Q(A)|,$$
where $A$ ranges over all subsets of the sample space.
\begin{property}[Property 2 \cite{schieler-journal}] \label{property-tv}
The total variation distance satisfies the following properties:
\begin{enumerate}[(a)]
%\item If $\Xcal$ is countable, then total variation can be rewritten as 
%\begin{equation}
%\lVert P - Q \rVert_{TV} = \frac12 \sum_{x\in\Xcal} |P(x)-Q(x)|.
%\end{equation}
\item Let $\eps>0$ and let $f(x)$ be a function in a bounded range with width $b \in\Rbb$. Then
\begin{equation}
\label{tvcontinuous}
\lVert P-Q \rVert_{TV} < \eps \:\Longrightarrow\: \big| \Ebb_P[f(X)] - \Ebb_Q[f(X)] \big | < \eps b.
\end{equation}
\item Total variation satisfies the triangle inequality. For any $R \in \Delta_{\Xcal}$, 
\begin{equation}
\lVert P - Q \rVert_{TV} \leq \lVert P - R \rVert_{TV} + \lVert R - Q \rVert_{TV}.
\end{equation}
\item Let $P_{X}P_{Y|X}$ and $Q_XP_{Y|X}$ be two joint distributions on $\Delta_{\Xcal\times\Ycal}$. Then 
\begin{equation}
\lVert P_XP_{Y|X} - Q_X P_{Y|X} \rVert_{TV} = \lVert P_X - Q_X \rVert_{TV}.
\end{equation}
\item For any $P,Q \in \Delta_{\Xcal\times\Ycal}$, 
\begin{equation}
\lVert P_X - Q_X \rVert_{TV} \leq \lVert P_{XY} - Q_{XY} \rVert_{TV}.
\end{equation}
\end{enumerate}
\end{property}

\subsection{The Likelihood Encoder}
We define the likelihood encoder, operating at rate $R$, which receives a sequence $x_1,...,x_n$ and maps it to a message $M \in [1:2^{nR}]$.  In normal usage, a decoder then uses $M$ to form an approximate reconstruction of the $x_1,...,x_n$ sequence.

The encoder is specified by a codebook of $y^n(m)$ sequences and a joint distribution $P_{XY}$.  Consider the likelihood function for each codeword, with respect to a memoryless channel from $Y$ to $X$, defined as follows:
$$\Lcal(m|x^n)\triangleq P_{X^n|Y^n}(x^n|y^n(m)).$$
A likelihood encoder is a stochastic encoder that determines the message index with probability proportional to $\Lcal(m|x^n)$, i.e. 
$$P_{M|X^n}(m|x^n)=\frac{\Lcal(m|x^n)}{\sum_{m'\in[1:2^{nR}]}\Lcal(m'|x^n)}\propto \Lcal(m|x^n).$$

\subsection{Soft-Covering Lemma}
Now we introduce the core lemma that serves as the foundation for this analysis. One can consider the role of the soft-covering lemma in analyzing the likelihood encoder as analogous to that of the J-AEP which is used for the analysis of joint-typicality encoders. The general idea of the soft-covering lemma is that the distribution induced by selecting uniformly from a random codebook and passing the codeword through a memoryless channel is close to an i.i.d. distribution as long as the codebook size is large enough.
\begin{lem}[Lemma 1.1 \cite{cuff-itw2013} and Lemma IV.1 \cite{cuff2012distributed}]
Given a joint distribution $P_{XY}$, let $\Ccal^{(n)}$ be a random collection of sequences $Y^n(m)$, with $m=1,...,2^{nR}$, each drawn independently and i.i.d. according to $P_Y$. Denote by $P_{X^n}$ the output distribution induced by selecting an index $m$ uniformly at random and applying $Y^n(m)$ to the memoryless channel specified by $P_{X|Y}$. Then if $R>I(X;Y)$,
$$\Ebb_{\cn}\lVert P_{X^n}-\prod_{t=1}^n P_X\rVert_{TV}\leq \epsilon_n\rightarrow_n 0.$$
\end{lem}

\subsection{Approximation Lemma}
\begin{lem} \label{helper}
For a distribution $P_{UVX}$ and $0<\eps<1$, if $\mathbb{P}[U\neq V]\leq \eps$, then 
$$\lVert P_{UX}-P_{VX}\rVert_{TV}\leq \eps.$$
\end{lem}
The proof is omitted due to a lack of space.

\section{Problem Setup and Result Review}
\subsection{Wyner-Ziv Model Review}
The source and side information $(X^n,B^n)$ is distributed i.i.d. according to $(X_t,B_t) \sim \Pbar_{XB}$. The system has the following constraints:
\begin{itemize}
\item Encoder $f_n: \mathcal{X}^n \mapsto \mathcal{M}$ (possibly stochastic).
\item Decoder $g_n: \mathcal{M}\times \mathcal{B}^n \mapsto \mathcal{Y}^n$ (possibly stochastic).
\item Compression rate: $R$, i.e. $|\mathcal{M}|=2^{nR}$.
\end{itemize}
The system performance is measured according to the following distortion metric: 
\begin{itemize}
\item Average distortion: $d(X^n, Y^n)=\frac1n\sum_{t=1}^n d(X_t,Y_t)$.\\
\end{itemize}
\vspace{-5mm}
\begin{defn}
A rate distortion pair $(R,D)$ is achievable if there exists a sequence of rate $R$ encoders and decoders $(f_n, g_n)$, such that $\Ebb \left[d(X^n,Y^n)\right]\leq_n D$.
\end{defn}
\begin{defn} \label{rate-distortion}
The rate distortion function is $R(D)\triangleq \inf_{\{(R,D) \text{ is achievable}\}} R$.
\end{defn}

\begin{figure}[htbp]
  \centering
  \includegraphics[width=7 cm]{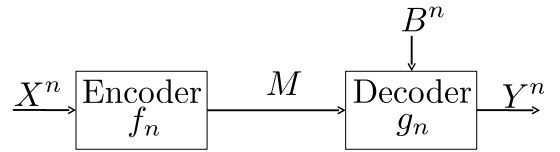}
\caption{The Wyner-Ziv problem: rate-distortion for source coding with side information at the decoder}
\label{setup_wz}
\end{figure}

The above mathematical formulation is illustrated in Fig. \ref{setup_wz}.
\vspace{-6mm}
\subsection{Rate-Distortion Function of Wyner-Ziv}
The solution to this source coding problem is given in \cite{wz}. The rate-distortion function with side information at the decoder is 
\begin{eqnarray}
R(D)&=& \min_{\Pbar_{V|XB}\in \mathcal{M}(D)} I_{\Pbar}(X;V|B), \label{rate}
\end{eqnarray}
where 
\begin{eqnarray}
\mathcal{M}(D)=\bigg\{\Pbar_{V|XB}: V\inout X\inout B, |\Vcal|\leq|\Xcal|+1, \nonumber \\\text{ and there exists }\nonumber\\
 \text{a function } \phi \text{ s.t. }
 \mathbb{E}\left[d(X,Y)\right]\leq D ,Y\triangleq \phi(V,B) \bigg\}. \label{md}
\end{eqnarray}

\section{Achievability Proof Using the Likelihood Encoder} \label{wz-proof}
Our proof technique involves using the likelihood encoder and a channel decoder and showing that the behavior of the system is approximated by a well-behaved distribution.  Exact bounds are obtained by using the soft-covering lemma to analyze how well the approximating distribution matches the system. For the readers' reference, a very short and simple achievability proof for point-to-point lossy compression was provided in \cite{cuff-itw2013}, which will serve to familiarize the reader with the proof techniques in this paper using the likelihood encoder.

We will introduce a virtual message which is produced by the encoder but not physically transmitted to the receiver so that this virtual message together with the actual message gives a high enough rate for applying the soft-covering lemma. Then we show that this virtual message can be reconstructed with vanishing error probability at the decoder by using the side information. This is analogous to the technique of random binning.

Let $R>R(D)$, where $R(D)$ is from (\ref{rate}). We prove that $R$ is achievable for distortion $D$. Let $M'$ be a virtual message with rate $R'$ which is not physically transmitted. By the rate-distortion formula $(\ref{rate})$, we can fix $\Pbar_{V|XB}\in\mathcal{M}(D)$, ($\Pbar_{V|XB}=\Pbar_{V|X}$) such that $R+R'>I_{\Pbar}(X;V)$ and $R'<I_{\Pbar}(V;B)$. We will use the likelihood encoder derived from $\Pbar_{XV}$ and a random codebook $\{v^n(m,m')\}$ generated according to $\Pbar_V$ to prove the result. The decoder will first use the transmitted message $M$ and the side information $B^n$ to decode $M'$ as $\hat{M}'$ and reproduce $v^n(M,\hat{M}')$. Then the reconstruction $Y^n$ is produced as a function of $B^n$ and $V^n$.

The distribution induced by the encoder and decoder is
\small
\begin{eqnarray}
\Pbf_{X^nB^nMM'\hat{M}'Y^n}\ \ \ \ \ \ \ \ \ \ \ \ \ \ \ \ \ \ \ \ \ \ \ \ \ \ \ \ \ \ \ \ \ \ \ \ \ \ \ \ \ \ \ \ \ \ \ \nonumber\\
\triangleq\Pbar_{X^nB^n}\Pbf_{MM'|X^n} \Pbf_{\hat{M}'|MB^n}\Pbf_{Y^n|M\hat{M}'B^n}\ \ \ \ \ \ \ \ \ \ \ \ \ \ \ \ \ \ \ \label{jointPP}\\
\triangleq\Pbar_{X^nB^n}\Pbf_{LE}(m,m'|x^n) \Pbf_D(\hat{m}'|m,b^n) \Pbf_\Phi (y^n|m,\hat{m}',b^n)\label{jointPP2}
\end{eqnarray}
\normalsize
where $\Pbf_{LE}$ is the likelihood encoder; $\Pbf_D(\hat{m}'|m,b^n)$ is the first part of the decoder that estimates $m'$ as $\hat{m}'$; and $\Pbf_\Phi(y^n|m,\hat{m}',b^n)$ is the second part of the decoder that reconstructs the source sequence. Note that the distributions are random due to the random codebook.

We now concisely restate the behavior of the encoder and decoder, as components of the induced distribution.

{\textbf{Codebook generation}}: We independently generate $2^{n(R+R')}$ sequences in $\mathcal{V}^n$ according to $\prod_{i=1}^n \Pbar_V(v_i)$ and index by $(m,m')\in[1:2^{nR}]\times[1:2^{nR'}]$. We use $\Ccal^{(n)}$ to denote the random codebook.

{\textbf{Encoder}}: The encoder $\Pbf_{LE}(m,m'|x^n)$ is the likelihood encoder that chooses $M$ and $M'$ stochastically with probability proportional to the likelihood function given by 
$$\Lcal(m,m'|x^n)=\Pbar_{X^n|V^n}(x^n|V^n(m,m')).$$

{\textbf{Decoder}}: The decoder has two steps. Let $\Pbf_D(\hat{m}'|m,b^n)$ be a good channel decoder (e.g. the maximum likelihood decoder) with respect to the sub-codebook ${\Ccal^{(n)}}(m)=\{v^n(m,a)\}_a$ and the memoryless channel $\Pbar_{B|V}$. For the second part of the decoder, let $\phi(\cdot,\cdot)$ be the function corresponding to the choice of $\Pbar_{V|XB}$ in $(\ref{md})$, that is $Y=\phi(V,B)$ and $\Ebb_{\Pbar}\left[d(X,Y)\right]\leq D$. Define $\phi^n(v^n, b^n)$ as the concatenation $\{\phi(v_t,b_t)\}_{t=1}^n$ and set the decoder $\Pbf_\Phi$ to be the deterministic function
$$\Pbf_\Phi(y^n|m,\hat{m}',b^n)\triangleq \mathbbm{1}\{y^n=\phi^n(V^n(m,\hat{m}'),b^n)\}.$$

{\textbf{Analysis:}} We will need three distributions for the analysis, the induced distribution $\Pbf$ and two approximating distributions $\Qbf^{(1)}$ and $\Qbf^{(2)}$. The idea is to show that 1) the system has nice behavior for distortion under $\Qbf^{(2)}$; and 2) $\Pbf$ and $\Qbf^{(2)}$ are close in total variation (averaged over the random codebook) through $\Qbf^{(1)}$. 
\begin{figure}[htbp]
  \centering
  \includegraphics[width=7.5 cm]{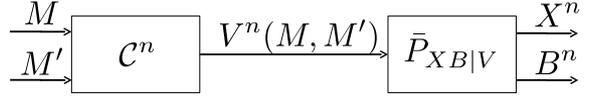}
\caption{Auxiliary distribution with test channel $\Pbar_{XB|V}$}
\label{auxiliary}
\end{figure}

Now we will design an auxiliary distribution $Q$ through a test channel as shown in Fig. \ref{auxiliary}. 
The joint distribution under $Q$ in Fig. \ref{auxiliary} can be written as
\small
\begin{eqnarray}
\mathbf{Q}_{X^nB^nV^nMM'}\ \ \ \ \ \ \ \ \ \ \ \ \ \ \ \ \ \ \ \ \ \ \ \ \ \ \ \ \ \ \ \ \ \ \ \ \ \ \ \ \ \ \ \ \ \ \ \ \ \ \ \ \ \ \nonumber\\
=Q_{MM'}\mathbf{Q}_{V^n|MM'}\mathbf{Q}_{X^nB^n|MM'}\ \ \ \ \ \ \ \ \ \ \ \ \ \ \ \ \ \ \ \ \ \ \ \ \ \ \ \ \ \ \ \ \ \ \ \ \nonumber\\
%&&\mathbf{Q}_{X^nB^nV^nMM'}\nonumber\\
%&=&Q_{MM'}\mathbf{Q}_{V^n|MM'}\mathbf{Q}_{X^nB^n|MM'}\nonumber
%\end{eqnarray}
%\begin{eqnarray}
=\frac{1}{2^{n(R+R')}}\mathbbm{1}\{v^n=V^n(m,m')\}\prod_{t=1}^n\Pbar_{XB|V}(x_t,b_t|V_t(m,m'))\nonumber\\
=\frac{1}{2^{n(R+R')}}\mathbbm{1}\{v^n=V^n(m,m')\}\prod_{t=1}^n\Pbar_{X|V}(x_t|v_t)\Pbar_{B|X}(b_t|x_t) \label{markovchain}
\end{eqnarray}
\normalsize
where $(\ref{markovchain})$ follows from the Markov chain under $\Pbar$, $V\inout X\inout B$. In fact, the reason for choosing the likelihood encoder lies in
\begin{eqnarray}
\mathbf{Q}_{MM'|X^n}= \mathbf{P}_{LE}. \label{enc}
\end{eqnarray}
Furthermore, it can be verified that
\begin{eqnarray}
\mathbb{E}_{\mathcal{C}^{(n)}}\left[\mathbf{Q}_{X^nB^nV^n}(x^n,b^n,v^n)\right]=\Pbar_{X^nB^nV^n}(x^n,b^n,v^n), \label{expectationQ}
\end{eqnarray}
where $\Pbar_{X^nB^nV^n}$ denotes the i.i.d. distribution $\prod_{t=1}^n\Pbar_{XBV}$.

Define two distributions $\Qbf^{(1)}$ and $\Qbf^{(2)}$ based on $\Qbf$ as follows:
\small
\begin{eqnarray}
%\mathbf{Q}^{(1)}_{X^nB^nV^nMM'\hat{M}'Y^n}(x^n,b^n,v^n,m,m',\hat{m}',y^n)\triangleq\nonumber\\
%\mathbf{Q}_{X^nB^nV^nMM'} P_D (\hat{m}'|m,b^n) P_\Phi (y^n|m,\hat{m}',b^n) \label{Q1}\\
%\mathbf{Q}^{(2)}_{X^nB^nV^nMM'\hat{M}'Y^n}(x^n,b^n,v^n,m,m',\hat{m}',y^n)\triangleq\nonumber\\
%\mathbf{Q}_{X^nB^nV^nMM'} P_D (\hat{m}'|m,b^n) P_\Phi (y^n|m,m',b^n) \label{Q2}
{\mathbf{Q}^{(1)}_{X^nB^nV^nMM'\hat{M}'Y^n}\triangleq\mathbf{Q}_{X^nB^nV^nMM'} \Pbf_D \Pbf_\Phi (y^n|m,\hat{m}',b^n)}\  \label{Q1}\\
{\mathbf{Q}^{(2)}_{X^nB^nV^nMM'\hat{M}'Y^n}\triangleq\mathbf{Q}_{X^nB^nV^nMM'} \Pbf_D \Pbf_\Phi (y^n|m,m',b^n)}. \label{Q2}
\end{eqnarray}
\normalsize

Notice that $\Qbf^{(2)}$ differs from $\Qbf^{(1)}$ by allowing the decoder to use $m'$ rather than $\hat{m}'$ when forming its reconstruction through $\phi^n$.

Therefore, on account of $(\ref{expectationQ})$, 
\small
\begin{eqnarray}
\mathbb{E}_{\mathcal{C}^{(n)}} \left[\mathbf{Q}^{(2)}_{X^nB^nV^nY^n}(x^n,b^n,v^n,y^n)\right]=\Pbar_{X^nB^nV^nY^n}(x^n,b^n,v^n,y^n).\nonumber
\end{eqnarray}
\normalsize
Consequently,
\begin{eqnarray}
\mathbb{E}_{\Ccal^{(n)}}\left[\mathbb{E}_{\mathbf{Q}^{(2)}}[d(X^n,Y^n)] \right]
= \mathbb{E}_{\Pbar}\left[d(X,Y)\right].\label{qq2}
\end{eqnarray}

%We set the source encoder in $(\ref{jointPP2})$ to be 
%\begin{eqnarray}
%\mathbf{P}_{MM'|X^n}\triangleq \mathbf{Q}_{MM'|X^n}, \label{enc}
%\end{eqnarray}
%i.e. the likelihood encoder that chooses $M$ and $M'$ stochastically with probability proportional to the likelihood function given by 
%$$\Lcal((m,m')|x^n)=\Pbar_{X^n|V^n}(x^n|V^n(m,m')).$$

Now applying the soft-covering lemma, since $R+R'>I_{\Pbar}(B,X;V)=I_{\Pbar}(X;V)$, we have
$$\mathbb{E}_{\mathcal{C}^{(n)}}\left[\lVert\Pbar_{X^nB^n}-\mathbf{Q}_{X^nB^n}\rVert_{TV}\right]\leq \epsilon_n\rightarrow_n 0.$$ 
And with $(\ref{jointPP2})$, $(\ref{enc})$, $(\ref{Q1})$, and Property \ref{property-tv}(c), we obtain
\begin{eqnarray}
\mathbb{E}_{\mathcal{C}^{(n)}}\left[\lVert \mathbf{P}_{X^nB^nMM'\hat{M}'Y^n}-\mathbf{Q}^{(1)}_{X^nB^nMM'\hat{M}'Y^n}\rVert_{TV}\right]
%&=&\mathbb{E}_{\mathcal{C}^n}\lVert\Pbar_{X^nB^n}-\mathbf{Q}_{X^nB^n}\rVert_{TV}\\
\leq \epsilon_n \label{PtoQ1}
\end{eqnarray}

Since by definition
$\mathbf{Q}^{(1)}_{X^nB^nMM'\hat{M}'}=\mathbf{Q}^{(2)}_{X^nB^nMM'\hat{M}'}$, 
$$\Upsilon\triangleq\Pbb_{\mathbf{Q}^{(1)}}[\hat{M}'\neq M']=\Pbb_{\mathbf{Q}^{(2)}}[\hat{M}'\neq M'].$$
Also, since $R'<I(V;B)$, the codebook is randomly generated, and $M'$ is uniformly distributed under $Q$, it is well known that the maximum likelihood decoder $\Pbf_D$ (as well as a variety of other decoders) will drive the error probability to zero as $n$ goes to infinity. Specifically,
$$\mathbb{E}_{\mathcal{C}^{(n)}}\left[\mathbb{P}_{\mathbf{Q}^{(1)}}[M'\neq \hat{M}']\right]\leq \delta_n\rightarrow_n 0.$$ 
Applying Lemma \ref{helper}, we obtain
\begin{eqnarray}
\Ebb_{\mathcal{C}^{(n)}} \lVert \mathbf{Q}^{(1)}_{X^nB^nM\hat{M}'}-\mathbf{Q}^{(2)}_{X^nB^nMM'} \rVert_{TV}
\leq\mathbb{E}_{\mathcal{C}^{(n)}}\left[\Upsilon\right]
\leq\delta_n.
\end{eqnarray}
Thus by Property \ref{property-tv}(c) and definitions $(\ref{Q1})$ and $(\ref{Q2})$,
\begin{eqnarray}
\Ebb_{\mathcal{C}^{(n)}} \left[\lVert \mathbf{Q}^{(1)}_{X^nB^nM\hat{M}'Y^n}-\mathbf{Q}^{(2)}_{X^nB^nMM'Y^n} \rVert_{TV}\right]\leq \delta_n. \label{Q1toQ2}
\end{eqnarray}
Combining $(\ref{PtoQ1})$ and $(\ref{Q1toQ2})$ and using Property \ref{property-tv}(b) (d), we have
\begin{eqnarray}
\Ebb_{\mathcal{C}^{(n)}}\left[ \lVert \mathbf{P}_{X^nY^n}-\mathbf{Q}^{(2)}_{X^nY^n}\rVert_{TV}\right]
%&\leq&\Ebb_{\mathcal{C}^n} \lVert \mathbf{P}_{X^nY^n}-\mathbf{Q}^{(1)}_{X^nY^n}\rVert_{TV}+\Ebb_{\mathcal{C}^n} \lVert \mathbf{Q}^{(1)}_{X^nY^n}-\mathbf{Q}^{(2)}_{X^nY^n}\rVert_{TV}\nonumber\\
\leq \epsilon_n+\delta_n,\label{PtoQ2}
\end{eqnarray}
where $\epsilon_n$ and $\delta_n$ are the error terms introduced from the soft-covering lemma and channel coding, respectively.

%The distortion under distribution $\mathbf{Q}^{(2)}$ averaged over the random codebook is given by the following. 
%Making use of $(\ref{expectationQ})$, it can be verified that the distortion under distribution $\mathbf{Q}^{(2)}$ averaged over the random codebook satisfies

%where $s=(x^n,v^n,b^n,m,m')$, $(\ref{m1})$ follows from $(\ref{expectationQ})$, and $(\ref{separable})$ follows from the definition of $\phi^n(\cdot,\cdot)$.

Using Property \ref{property-tv}(a)  and $(\ref{qq2})$ and $(\ref{PtoQ2})$, we have
\begin{eqnarray}
\Ebb_{\Ccal^{(n)}}\left[ \Ebb_{\mathbf{P}}[d(X^n,Y^n)]\right]
%&\leq&\Ebb_{\mathcal{C}^n} \Ebb_{\mathbf{Q}^{(2)}}[d(X^n,Y^n)]+d_{max}(\epsilon_n+\delta_n)\\
\leq\Ebb_{\Pbar}\left[d(X,Y)\right]+d_{max}(\epsilon_n+\delta_n). \label{endp}
\end{eqnarray}
Therefore, there exists a codebook under which
$$\Ebb_{P}[d(X^n,Y^n)]\leq_n D.$$

\section{Extension to Distributed Lossy Source Compression}
The application of the likelihood encoder can go beyond single-user communications. In this section, we will outline an alternative proof for achieving the Berger-Tung inner bound.

\subsection{Berger-Tung Model Review}
We now assume a pair of correlated sources $({X_1}^n,{X_2}^n)$, distributed i.i.d. according to $({{X_1}}_t,{X_2}_t)\sim \Pbar_{{X_1}{X_2}}$, independent encoders, and a joint decoder, satisfying the following constraints:
\begin{itemize}
\item Encoder 1 ${f_1}_n: {\mathcal{X}_1}^n \mapsto \mathcal{M}_1$ (possibly stochastic).
\item Encoder 2 ${f_2}_n: {\mathcal{X}_2}^n \mapsto \mathcal{M}_2$ (possibly stochastic).
\item Decoder $g_n: \mathcal{M}_1\times \mathcal{M}_2 \mapsto {\mathcal{Y}_1}^n\times {\mathcal{Y}_2}^n$ (possibly stochastic).
\item Compression rates: $R_1, R_2$, i.e. $|\mathcal{M}_1|=2^{nR_1}$, $|\mathcal{M}_2|=2^{nR_2}$.
\end{itemize}
The system performance is measured according to the following distortion metric: 
\begin{itemize}
\item $\mathbb{E}[d_k({X_k}^n, {Y_k}^n)] =\frac1n\sum_{t=1}^nd_k({{X_k}_t,{Y_k}_t})$, $k=1,2$, where $d_k(\cdot,\cdot)$ can be different distortion measures for different $k$.
\end{itemize}
\begin{defn}
$(R_1,R_2)$ is achievable under distortion level $(D_1,D_2)$ if there exists a sequence of rate $(R_1,R_2)$ encoders and decoders $({f_1}_n, {f_2}_n, g_n)$ such that
$$\mathbb{E}[d_1({X_1}^n, {Y_1}^n)] \leq_n D_1,$$
$$\mathbb{E}[d_2({X_2}^n, {Y_2}^n)] \leq_n D_2.$$
\end{defn}

%The closure of the optimal rate region is not yet known in general. But an inner bound, reproduced below, was given in \cite{tung} and \cite{berger1977}, and is known as the Berger-Tung inner bound:
%\begin{eqnarray}
%R_1&\geq& I_{\Pbar}(X_1;U_1|U_2), \label{rate1}\\
%R_2&\geq& I_{\Pbar}(X_2;U_2|U_1), \label{rate2}\\
%R_1+R_2&\geq& I_{\Pbar}(X_1,X_2;U_1,U_2) \label{rate12}
%\end{eqnarray}
%for some $\Pbar_{U_1X_1X_2U_2}=\Pbar_{X_1X_2}\Pbar_{U_1|X_1}\Pbar_{U_2|X_2}$, and functions $\widehat{x_k}(\cdot,\cdot)$ such that $\mathbb{E}[d_k(X_k,Y_k)]\leq D_k$, where $Y_k\triangleq \widehat{x_k}(U_1,U_2), k=1,2$.

The achievable rate region is not yet known in general. But an inner bound, reproduced below, was given in \cite{tung} and \cite{berger1977} and is known as the Berger-Tung inner bound. The rates $(R_1,R_2)$ are achievable if
\begin{eqnarray}
R_1&>& I_{\Pbar}(X_1;U_1|U_2), \label{rate1}\\
R_2&>& I_{\Pbar}(X_2;U_2|U_1), \label{rate2}\\
R_1+R_2&>& I_{\Pbar}(X_1,X_2;U_1,U_2) \label{rate12}
\end{eqnarray}
for some $\Pbar_{U_1X_1X_2U_2}=\Pbar_{X_1X_2}\Pbar_{U_1|X_1}\Pbar_{U_2|X_2}$, and functions $\phi_k(\cdot,\cdot)$ such that $\mathbb{E}[d_k(X_k,Y_k)]\leq D_k$, where $Y_k\triangleq \phi_k(U_1,U_2), k=1,2$. \footnote{This region, after optimizing over auxiliary variables, is in fact not convex, so it can be improved to the convex hull through time-sharing.}

\subsection{Proof Sketch Using the Likelihood Encoder}
%For simplicity, we will focus on the corner points, $C_1\triangleq \left(I_{\Pbar}(X_1;U_1),I_{\Pbar}(X_2;U_2|U_1)\right)$ and $C_2\triangleq \left(I_{\Pbar}(X_1;U_1|U_2),I_{\Pbar}(X_2;U_2)\right)$, of the region given in $(\ref{rate1})$ through $(\ref{rate12})$ and use the time-sharing argument to get the complete known region. Below we demonstrate how to achieve $C_1$. By symmetry, $C_2$ can be achieved in a similar fashion by interchanging the roles of the variables.
%
%Fix a $\Pbar_{U_1U_2|X_1X_2}=\Pbar_{U_1|X_1}\Pbar_{U_2|X_2}$ and functions $\widehat{x_k}(\cdot,\cdot)$ such that $Y_k=\widehat{x_k}(U_1,U_2)$ and $\Ebb_{\Pbar}\left[d_k(X_k,Y_k)\right]<D_k$.  Note that $U_1\inout X_1\inout X_2\inout U_2$ forms a Markov chain under $\Pbar$. We choose the rates to satisfy: $R_1>I_{\Pbar}(X_1;U_1)$, $R_2'<I_{\Pbar}(U_1;U_2)$, $R_2+R_2'>I_{\Pbar}(X_2;U_2)$ and $R_1+R_2>I_{\Pbar}(X_1,X_2;U_1,U_2)$.
For simplicity, we will focus on the corner points, $C_1\triangleq \left(I_{\Pbar}(X_1;U_1),I_{\Pbar}(X_2;U_2|U_1)\right)$ and $C_2\triangleq \left(I_{\Pbar}(X_1;U_1|U_2),I_{\Pbar}(X_2;U_2)\right)$, of the region given in $(\ref{rate1})$ through $(\ref{rate12})$ and use convexity to claim the complete region. Below we demonstrate how to achieve $C_1$. The point $C_2$ follows by symmetry.

Fix a $\Pbar_{U_1U_2|X_1X_2}=\Pbar_{U_1|X_1}\Pbar_{U_2|X_2}$ and functions ${\phi_k}(\cdot,\cdot)$ such that $Y_k={\phi_k}(U_1,U_2)$ and $\Ebb_{\Pbar}\left[d_k(X_k,Y_k)\right]<D_k$.  Note that $U_1\inout X_1\inout X_2\inout U_2$ forms a Markov chain under $\Pbar$. We must show that any rates $(R_1,R_2)$ satisfying $R_1>I_{\Pbar}(X_1;U_1)$ and $R_2>I_{\Pbar}(X_2;U_2|U_1)$ are achievable.

First we will use the likelihood encoder derived from $\Pbar_{X_1U_1}$ and a random codebook $\{{u_1}^n(m_1)\}$ generated according to $\Pbar_{U_1}$ for Encoder 1. Then we will use the likelihood encoder derived from $\Pbar_{X_2U_2}$ and another random codebook $\{{u_2}^n(m_2,m_2')\}$ generated according to $\Pbar_{U_2}$ for Encoder 2. The decoder will use the transmitted message $M_1$ to decode ${U_1}^n$, as in the point-to-point case,  and use the transmitted message $M_2$ along with the decoded ${U_1}^n$ to decode $M_2'$ as $\hat{M}_2'$, as in the Wyner-Ziv case, and reproduce $u_2^n(M_2,\hat{M}_2')$. Finally, the decoder outputs the reconstructions ${Y_k}^n$ as functions of ${U_1}^n$ and ${U_2}^n$.

The distribution induced by the encoders and decoder is 
$$\Pbf_{{X_1}^n{X_2}^n{U_1}^nM_1M_2M_2'\hat{M}_2'{Y_1}^n{Y_2}^n}=\Pbar_{{X_1}^n{X_2}^n}\Pbf_1\Pbf_2$$
\begin{eqnarray}
\Pbf_1&\triangleq& \Pbf_{M_1|{X_1}^n}\Pbf_{{U_1}^n|M_1}\\
\Pbf_2&\triangleq& \Pbf_{M_2M_2'|{X_2}^n}\Pbf_{\hat{M}_2'|M_2{U_1}^n}\prod_{k=1,2}\Pbf_{{Y_k}^n|{U_1}^nM_2\hat{M}_2'}\\
&\triangleq& \Pbf_{M_2M_2'|{X_2}^n} \Pbf_{D} \prod_{k=1,2} \Pbf_{\Phi, k}, \label{defp2}
\end{eqnarray}
where again $M_2'$ plays the role of the virtual message that is not physically transmitted as in the Wyner-Ziv case.

{\textbf{Codebook generation}}: We independently generate $2^{nR_1}$ sequences in ${\mathcal{U}_1}^n$ according to $\prod_{t=1}^n\Pbar_{U_1}({u_1}_t)$ and index them by $m_1\in[1:2^{nR_1}]$, and independently generate $2^{n(R_2+R_2')}$ sequences in ${\mathcal{U}_2}^n$ according to $\prod_{t=1}^n\Pbar_{U_2}({u_2}_t)$ and index them by $(m_2,m_2')\in[1:2^{nR_2}]\times[1:2^{nR_2'}]$. We use $\Ccal_1^{(n)}$ and $\Ccal_2^{(n)}$ to denote the two random codebooks, respectively. 

{\textbf{Encoders}}: Encoder 1 $\Pbf_{M_1|{X_1}^n}$ is the likelihood encoder according to $\Pbar_{{X_1}^n{U_1}^n}$ and $\Ccal_1^{(n)}$. Encoder 2 $\Pbf_{M_2M_2'|{X_2}^n}$ is the likelihood encoder according to $\Pbar_{{X_2}^n{U_2}^n}$ and $\Ccal_2^{(n)}$. 

{\textbf{Decoder}}:  First, let $\Pbf_{{U_1}|M_1}$ be a $\Ccal_1^{(n)}$ codeword lookup decoder. Then, let $\Pbf_D(\hat{m}_2'|m_2,{u_1}^n)$ be a good channel decoder with respect to the sub-codebook $\Ccal_2^{(n)}(m_2)=\{{u_2}^n(m_2,a)\}_a$ and the memoryless channel $\Pbar_{U_1|U_2}$. Last, define ${\phi_k}^n({u_1}^n,{u_2}^n)$ as the concatenation $\{{\phi_k}({u_1}_t,{u_2}_t)\}_{t=1}^n$ and set the decoders $\Pbf_{\Phi,k}$ to be the deterministic functions
$$\Pbf_{\Phi,k}({y_k}^n|{u_1}^n,m_2,\hat{m}_2')\triangleq\mathbbm{1}\{{y_k}^n={{\phi_k}^n({u_1}^n,{U_2}^n(m_2,\hat{m}_2'))}\}.$$

{\textbf{Analysis}}: We will need the following distributions: the induced distribution $\Pbf$ and auxiliary distributions $\Qbf_1$ and $\Qbf_1^*$. The general idea of the proof is as follows: Encoder 1 makes $\Pbf$ and $\Qbf_1$ close in total variation. Distribution $\Qbf_1^*$ (random only with respect to the second codebook $\Ccal_2^{(n)}$) is the expectation of $\Qbf_1$ over the random codebook $\Ccal_1^{(n)}$. This is really the key step in the proof. By considering the expectation of the distribution with respect to $\Ccal_1^{(n)}$, we effectively remove Encoder 1 from the problem and turn the message from Encoder 1 into memoryless side information at the decoder. Hence, the two distortions (averaged over $\Ccal_1^{(n)}$) under $\Pbf$ are roughly the same as the distortions under $\Qbf_1^*$, which is a much simpler distribution. %Note that the lossy representation ${U_1}^n$ of ${X_1}^n$ from the first stage serves as the side information at the decoder for the second stage. Like the Wyner-Ziv model, the system has nice behavior in terms of distortion under $\Qbf_2^{(2)}$; $\Qbf_1^*$ and $\Qbf_2^{(2)}$ are close in total variation (averaged over $\Ccal_2^{(n)}$) through $\Qbf_2^{(1)}$. Hence, the distortions under $\Qbf_1^*$ and $\Qbf_2^{(2)}$ are roughly the same (averaged over $\Ccal_2^{(n)}$). Finally, we average over both $\Ccal_1^{(n)}$ and $\Ccal_2^{(n)}$ to obtain that the distortions under $\Pbf$ and $\Qbf_2^{(2)}$ are close. Below we outline the technical details. 
We then recognize $\Qbf_1^*$ as precisely $\Pbf$ in $(\ref{jointPP2})$ from the Wyner-Ziv proof of the previous section, with a source pair $(X_1,X_2)$, a pair of reconstructions $(Y_1,Y_2)$ and $U_1$ as the side information.

1) The auxiliary distribution $\Qbf_1$ takes the following form:
$${\mathbf{Q}_1}_{{X_1}^n{X_2}^n{U_1}^nM_1M_2M_2'\hat{M}_2'{Y_1}^n{Y_2}^n}={\mathbf{Q}_1}_{M_1{U_1}^n{X_1}^n {X_2}^n}\Pbf_2$$
\begin{eqnarray}
&&{\mathbf{Q}_1}_{M_1{U_1}^n{X_1}^n {X_2}^n}(m_1,{u_1}^n,{x_1}^n,{x_2}^n)\nonumber\\
&=&\frac{1}{2^{nR_1}}\mathbbm{1}\{{u_1}^n={U_1}^n(m_1)\}\Pbar_{{X_1}^n|{U_1}^n}({x_1}^n|{u_1}^n)\nonumber\\
&&\Pbar_{{X_2}^n|{X_1}^n}({x_2}^n|{x_1}^n)\label{qqq}
\end{eqnarray}
where $\Pbf_2$ was defined earlier in $(\ref{defp2})$.
Applying the soft-covering lemma, since $R_1>I_{\Pbar}(X_1;U_1)$, 
$$\mathbb{E}_{\Ccal_1^{(n)}}\left[\lVert{\mathbf{Q}_1}_{X_1^n}-\Pbar_{X_1^n}\rVert_{TV}\right]\leq {\epsilon_1}_n\rightarrow_n 0.$$
Consequently,
\begin{eqnarray}
\mathbb{E}_{\Ccal_1^{(n)}}\left[\lVert {\mathbf{Q}_1}-{\mathbf{P}} \rVert_{TV}\right]\leq {\epsilon_1}_n, \label{tv-layer1}
\end{eqnarray}
where $\Qbf_1$ and $\Pbf$ are distributions over random variables ${X_1}^n,{X_2}^n,{U_1}^n,M_1,M_2,M_2',\hat{M}_2',{Y_1}^n,$ and ${Y_2}^n$.

2) Taking the expectation over codebook $\Ccal_1^{(n)}$, we define
\begin{eqnarray}
&&{\Qbf^*_1}_{{X_1}^n{X_2}^n{U_1}^nM_2M_2'\hat{M}_2'{Y_1}^n{Y_2}^n}\nonumber\\
&\triangleq& \mathbb{E}_{\Ccal_1^{(n)}}\left[{\mathbf{Q}_1}_{{X_1}^n{X_2}^n{U_1}^nM_2M_2'\hat{M}_2'{Y_1}^n{Y_2}^n}\right].
\end{eqnarray}

Note that under this definition of $\Qbf^*_1$, we have
\small
\begin{eqnarray}
{\Qbf^*_1}_{{X_1}^n{X_2}^n{U_1}^nM_2M_2'\hat{M}_2'{Y_1}^n{Y_2}^n}({x_1}^n,{x_2}^n,{u_1}^n,m_2,m_2',\hat{m}_2',{y_1}^n,{y_2}^n)\nonumber\\
=\Pbar_{{X_1}^n{X_2}^n{U_1}^n}({x_1}^n,{x_2}^n,{u_1}^n)\Pbf_2(m_2,m_2',\hat{m}_2',{y_1}^n,{y_2}^n|{x_2}^n,{u_1}^n).\nonumber
\end{eqnarray}

\normalsize
By Property \ref{property-tv}(b),
\begin{eqnarray}
&&\mathbb{E}_{\Ccal_1^{(n)}}\left[ \Ebb_{\mathbf{P}}\left[d_k({X_k}^n,{Y_k}^n)\right]\right]\nonumber\\
&\leq&\mathbb{E}_{\Ccal_1^{(n)}} \left[\Ebb_{\mathbf{Q}_1}[d_k({X_k}^n,{Y_k}^n)]\right]+d_{max}{\epsilon_1}_n\\
&=&\Ebb_{\Qbf_1^*}\left[d_k({X_k}^n,{Y_k}^n)\right]+d_{max}{\epsilon_1}_n. \label{rr1}
\end{eqnarray}

Note that $\Qbf^*_1$ is exactly of the form of the induced distribution $\Pbf$ in the Wyner-Ziv proof of the previous section, with the inconsequential modification that there are two reconstructions and two distortion functions. With the same techniques as $(\ref{Q1})$ through $(\ref{endp})$, we obtain 
\begin{eqnarray}
&&\Ebb_{\Ccal_2^{(n)}} \left[\Ebb_{\mathbf{Q}^*_1}\left[d_k({X_k}^n,{Y_k}^n)\right]\right] \nonumber\\
&\leq& \Ebb_{\Pbar}\left[d_k(X_k,Y_k)\right]+d_{max} ({\epsilon_2}_n+\delta_n),\label{DQstar}
\end{eqnarray}
where ${\epsilon_2}_n$ and $\delta_n$ are error terms introduced from the soft-covering lemma and channel decoding, respectively.

Finally, taking the expectation over $\Ccal_{1}^{(n)}$ and using $(\ref{rr1})$ and $(\ref{DQstar})$,
$$\mathbb{E}_{\Ccal_2^{(n)}}\left[\mathbb{E}_{\Ccal_1^{(n)}}\left[\Ebb_{\mathbf{P}}\left[d_k({X_k}^n,{Y_k}^n)\right]\right]\right]\leq D_k+d_{max}({\epsilon_1}_n+{\epsilon_2}_n+\delta_n).$$

\bibliographystyle{ieeetr}

\bibliography{likelihood_encoder_final}

\begin{thebibliography}{10}

\bibitem{shannon-math}
C.~E. Shannon, ``A mathematical theory of communication,'' {\em Bell Sys. Tech.
  Journal}, vol.~27, pp.~379--423, 623--656, 1948.

\bibitem{shannon-rd}
C.~E. Shannon, ``Coding theorems for a discrete source with a fidelity
  criterion,'' {\em IRE National Convention Record, Part 4}, pp.~142--163,
  1959.

\bibitem{wz}
A.~Wyner and J.~Ziv, ``The rate-distortion function for source coding with side
  information at the decoder,'' {\em IEEE Transactions on Information Theory},
  vol.~22, no.~1, pp.~1--10, 1976.

\bibitem{tung}
S.-Y. Tung, {\em Multiterminal Source Coding}.
\newblock PhD thesis, Cornell University, Ithaca, NY, May, 1978.

\bibitem{berger1977}
T.~Berger, ``Multiterminal source coding,'' {\em The Information Theory
  Approach to Communications}, vol.~229, pp.~171--231, 1977.

\bibitem{berger1989}
T.~Berger and R.~W. Yeung, ``Multiterminal source encoding with one distortion
  criterion,'' {\em IEEE Transactions on Information Theory}, vol.~35, no.~2,
  pp.~228--236, 1989.

\bibitem{cover}
T.~M. Cover and J.~A. Thomas, {\em Elements of Information Theory}.
\newblock John Wiley \& Sons, 2012.

\bibitem{network-it}
A.~El~Gamal and Y.-H. Kim, {\em Network Information Theory}.
\newblock Cambridge University Press, 2011.

\bibitem{hybrid}
P.~Minero, S.~H. Lim, and Y.-H. Kim, ``Hybrid coding: An interface for joint
  source-channel coding and network communication,'' {\em arXiv preprint
  arXiv:1306.0530}, 2013.

\bibitem{lapidoth}
A.~Lapidoth and S.~Tinguely, ``Sending a bivariate gaussian over a gaussian
  mac,'' {\em IEEE Transactions on Information Theory}, vol.~56,
  pp.~2714--2752, June 2010.

\bibitem{cuff-itw2013}
P.~Cuff and E.~C. Song, ``The likelihood encoder for source coding,'' in {\em
  Proc. IEEE Information Theory Workshop (ITW)}, 2013.

\bibitem{cuff2012distributed}
P.~Cuff, ``Distributed channel synthesis,'' {\em IEEE Transactions on
  Information Theory}, vol.~59, no.~11, pp.~7071--7096, 2013.

\bibitem{cuff-permuter}
P.~W. Cuff, H.~H. Permuter, and T.~M. Cover, ``Coordination capacity,'' {\em
  IEEE Transactions on Information Theory}, vol.~56, no.~9, pp.~4181--4206,
  2010.

\bibitem{schieler-journal}
C.~Schieler and P.~Cuff, ``Rate-distortion theory for secrecy systems,'' {\em
  CoRR}, vol.~abs/1305.3905, 2013.

\bibitem{wyner}
A.~D. Wyner, ``The common information of two dependent random variables,'' {\em
  IEEE Transactions on Information Theory}, vol.~21, no.~2, pp.~163--179, 1975.

\bibitem{han-verdu}
T.~Han and S.~Verd\'{u}, ``Approximation theory of output statistics,'' {\em
  IEEE Transactions on Information Theory}, vol.~39, no.~3, pp.~752--772, 1993.

\bibitem{coord}
P.~Cuff, H.~Permuter, and T.~Cover, ``Coordination capacity,'' {\em IEEE
  Transactions on Information Theory}, vol.~56, pp.~4181--4206, Sept 2010.

\bibitem{jeon}
J.~Jeon, ``A generalized typicality for abstract alphabets,'' {\em arXiv
  preprint arXiv:1401.6728}, 2014.

\end{thebibliography}
\end{document}